\documentclass[aps,prl,twocolumn,groupedaddress]{revtex4}

\usepackage{graphicx}
\usepackage{dcolumn}
\usepackage{bm}

\begin{document}

\preprint{APS/123-QED}

\title{Microsecond resolution of quasiparticle tunneling in the single-Cooper-pair-transistor}

\author{A. J. Ferguson}
 \email{andrew.ferguson@unsw.edu.au}
\author{N. A. Court}
\author{F. E. Hudson}
\author{R. G. Clark}%
\affiliation{%
Australian Research Centre of Excellence for Quantum Computer
Technology, University of New South Wales, Sydney NSW 2052,
Australia
}%

\date{\today}

\begin{abstract}
We present radio-frequency measurements on a
single-Cooper-pair-transistor in which individual quasiparticle
poisoning events were observed with microsecond temporal resolution.
Thermal activation of the quasiparticle dynamics is investigated,
and consequently, we are able to determine energetics of the
poisoning and un-poisoning processes. In particular, we are able to
assign an effective quasiparticle temperature to parameterize the
poisoning rate.

\end{abstract}

\pacs{Valid PACS appear here}
\maketitle Operation of both the single-Cooper-pair-transistor
(SCPT) and the Cooper-pair-box (CPB) rely on the coherent tunneling
of a single Cooper-pair between a reservoir and a tunnel-coupled
island. This coherent phenomenon is the basis of CPB charge qubits
\cite{nak99,wal04} and low-dissipation electrometry using the SCPT
\cite{zor01,man04}. One of the challenges that faces these devices
is avoiding the incoherent tunneling of quasiparticles, often
referred to as quasiparticle poisoning. The effect of quasiparticle
poisoning is to change the charge on the device island by an
electron and halt the coherent tunneling of Cooper-pairs. This is
especially undesirable for the CPB qubit where it can be a major
source of decoherence \cite{lut05,lut06}.

Quasiparticle poisoning has been extensively studied with a wide
variation in behavior observed \cite{man04,joy94,eil94,tuo97,ama94}.
However, a model suggested by Aumentado \textit{et al.} appears to
successfully explain the phenomenon \cite{aum04}. In this model
there is some un-known (and possibly non-equilibrium) source of
quasiparticles in the device leads. These quasiparticles are able to
tunnel onto (poisoning) the device island which usually acts as a
quasiparticle trap. Subsequently the trapped quasiparticle is
thermally excited (un-poisoning) out of the trap, and the island
returns to its even-state.

\begin{figure}
\includegraphics[width=7.0cm]{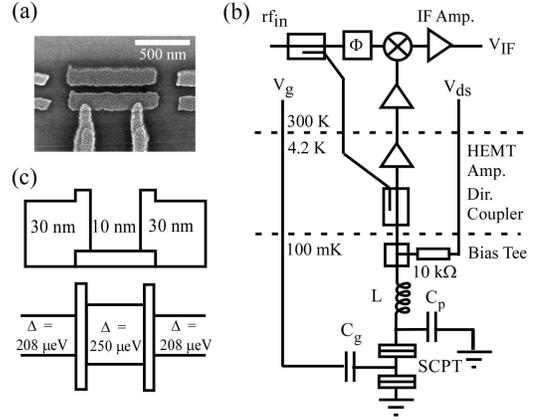}\\
\caption{(a) Scanning electron micrograph of the device. (b)
Simplified rf circuit diagram. The rf-carrier signal reflected from
the tank circuit is amplified by a cryogenic amplifier with a gain
of 38 dB at 4.2 K. After a further 30 dB of amplification at room
temperature, the carrier is phase shifted and homodyne detected. (c)
Profile of the SCPT showing aluminum film thickness and the
resulting change in $\Delta$.}\label{}
\end{figure}

While most previous investigations of quasiparticle poisoning have
been performed with a relatively low bandwidth, we note a very
recent careful study of oxygen-doped aluminium SCPTs measured by an
rf-technique sensitive to the Josephson inductance \cite{naa06}. In
that case, detailed measurements of the temperature dependence of
the poisoned state lifetime allowed determination of a quasiparticle
trap depth on the island. In this paper we present measurements of a
SCPT, made by a different technique, embedded in a related
radio-frequency (rf) tank circuit. A temperature dependent study
allowed the energetics of both the poisoning and un-poisoning
processes to be determined. In particular, a measurement of the
thermal activation of the un-poisoned state lifetime enabled an
effective quasiparticle temperature to be deduced which is an
experimentally useful way to parameterize quasiparticle poisoning.

We engineer the SCPT to have a greater superconducting gap
($2\Delta$) for the island than the leads $\Delta_i>\Delta_l$
\cite{aum04} by making use of the rapid enhancement of $\Delta$ with
decreasing film thickness \cite{mes71,gun04} (fig. 1(c)). This
reduces the depth of the quasiparticle trap, allowing quasiparticles
to be more easily thermally excited out from the island. The island
is made from a 10 nm thick film ($\Delta_i=250\pm15$ $\mu$eV) while
the leads have a thickness of 30 nm ($\Delta_l=208\pm10$ $\mu$eV),
with $\Delta$ determined by measuring the onset of quasiparticle
tunneling in SIS junctions. The device pattern (fig. 1(a)) is
defined in polymer bilayer resist by electron beam lithography and
the aluminum thermally evaporated at a rate of 0.1 nms$^{-1}$ onto a
liquid nitrogen cooled stage. With this technique we were able to
achieve electrically continuous films down to a thickness of 5 nm
\cite{fer06}. A controlled oxidization (35 mTorr for 2 minutes)
between the evaporations defines the tunnel barriers.

\begin{figure}
\includegraphics[width=8.0cm]{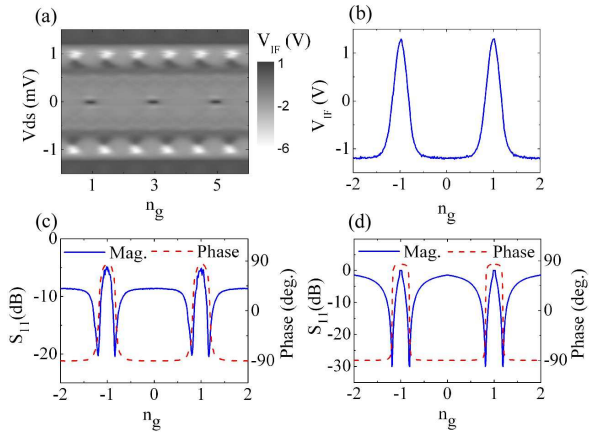}\\
\caption{(a) Coulomb diamonds showing the 2e-periodic supercurrent
at zero-bias and e-periodic transport at finite bias. (b) Amplified
mixer output for $V_{ds}=0$ showing the form of the supercurrent
oscillations.  (c) Network analyzer measurement (heavily averaged)
of magnitude and phase of rf-carrier across the supercurrent
oscillations. The incident power is -107 dBm. (d) Model of the
amplitude and phase response across the supercurrent
oscillations.}\label{}
\end{figure}

The circuit used for this experiment is the same as commonly used
for the rf-SET (fig. 1(b)) \cite{sch98,ros04}. A resonant circuit at
$326$ MHz is formed by a chip inductor ($L=470$ nH), a parasitic
capacitance ($C_p=0.51$ pF) and the SCPT. The reflection coefficient
($S_{11}=|Z_T-50/Z_T+50|$) of a small ($\sim\mu$V) incident
rf-carrier signal at the circuit resonance ($\omega^2=1/LC$) is
determined by mis-match of the tank circuit impedance  ($Z_T=L/RC$)
to a 50 $\Omega$ co-axial cable. The reflected carrier signal is
then amplified by a low-noise cryogenic amplifier. Following further
amplification at room temperature, the rf-carrier is demodulated by
mixing with a local oscillator at the carrier frequency (a technique
sensitive to both phase and amplitude of the reflected rf-carrier).
The resulting intermediate frequency (IF) output is further
amplified and recorded on an oscilloscope.

When Coulomb diamonds are measured a 2-e periodic supercurrent is
observed at zero bias along with e-periodic features at finite bias
due to a combination of Josephson quasiparticle (JQP) resonances and
Coulomb blockade of quasiparticle tunneling (fig. 2(a)). Similar
behavior was seen in a number of other devices. The 4.2 K resistance
of this device was 47 k$\Omega$ and the charging energy
$E_c=e^2/2C_\Sigma=77$ $\mu$eV, as determined from normal-state
Coulomb diamonds measured at $B=2.5$ T. Estimating the Josephson
energy per junction from the 4.2 K resistance and the
Ambegoakar-Baratoff relation ($E_J\sim
\frac{h\Delta_i\Delta_l}{4(\Delta_i+\Delta_l)e^2R}$) we find
$E_J=33$ $\mu$eV.

Taking a single trace over the supercurrent oscillations at
$V_{ds}=0$ (fig. 2(b)), a change in polarity of the mixer output
occurs indicating a phase shift of the reflected rf-carrier. Further
investigation with a network analyzer (fig. 2(c)), shows large
changes in both the amplitude and phase across the supercurrent
oscillations. For the amplitude component, there is a high
reflection coefficient at both $n_g=1$ (on supercurrent maxima) and
$n_g=0$, and a minima in reflection coefficient on the sides of the
supercurrent oscillation. There is also a large phase-shift
($\delta\theta=178$ degrees) between $n_g=1$ and $n_g=0$, with the
phase shifts coinciding with the amplitude minima.

To understand the behavior of this circuit we develop a model in
which the rf response depends on the ratio of the driving current
($I_{rf}\sim V_{rf}/\omega L=1$ nA at -107 dBm) to the switching
current ($I_{sw}$) of the SCPT. We calculate switching currents
using a 2-band model of the SCPT \cite{joy94} finding a maximum
$I_{sw}=4.6$ nA for the ground-band at $n_g=1$. If $I_{rf}<I_{sw}$,
we assume the SCPT remains in current-mode and presents zero
resistance. This is impedance transformed by the tank circuit to
yield $Z_T\sim\infty$ and causes almost complete reflection.

In the case where $I_{rf}>I_{sw}$, we assume a hysteresis loop at
the carrier frequency in which the device is partly (for
$I_{rf}(t)<I_{sw}$) in current mode and partly in voltage mode (for
$I_{rf}(t)>I_{sw}$). This leads to an average resistance $\langle
R\rangle$ which is transformed by the tank circuit to $Z_T=L/\langle
R\rangle C$. Using $I_{sw}$ from the 2-band model, an expression for
$\langle R\rangle$ \cite{eq} and the tank circuit parameters, we
simulate the amplitude and phase response of the device (fig. 2(d)).
We note a reduced value of $E_J=8$ $\mu$eV was taken to account for
the suppression of $I_{sw}$ due to environmental effects. The
amplitude and phase response are well-modeled, with the large phase
shift occurring as the resonant circuit changes between under
($n_g=0,\langle R\rangle
>1$ M$\Omega$) and over-damping ($n_g=1,\langle R\rangle=0$
$\Omega$). A phase shift is expected in the SCPT due to the
Josephson inductance \cite{sil04}, however in this case it can be
explained by the resonant circuit going through critical damping.

Charge sensitivity is determined by applying a 1 MHz gate signal of
0.026 e rms and measuring the signal to noise ratio (SNR) of the
resulting sidebands with a spectrum analyzer. Since there is a phase
component to our signal we perform this measurement after
de-modulation. Using the formula $\Delta
q_{rms}\times10^{-SNR/20}/\sqrt{2B}$, where B is resolution
bandwidth, we find a sensitivity of $1.5\times10^{-5 }$ $eHz^{-0.5}$
which is comparable to superconducting and normal state rf-SETs
\cite{ros04}.

Two-level switching behavior occurs with the device biased on a
supercurrent peak and $V_{IF}$ monitored as a function of time. The
inset in fig. 3(a) shows a representative 4 ms time record. The
positive level at $V_{IF}=1.6$ V corresponds to the the top of the
supercurrent peak while the negative level $V_{IF}=-1.0$ V
corresponds to the signal in the trough between peaks. We attribute
the positive voltage state to an even (un-poisoned) state where
there are only Cooper pairs on the island. By contrast, negative
voltages correspond to a 'poisoned-state' where a single
quasiparticle occupies the island and the supercurrent peak is
shifted from $n_g=1$ to $n_g=0$. We record traces of length 400 ms,
consisting of $10^6$ data points, in order to obtain reliable
statistics.

\begin{figure}
\includegraphics[width=6.8cm]{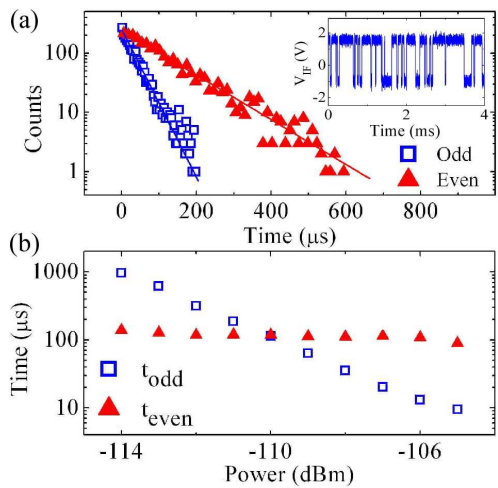}\\
\caption{(a) A histogram of times spent in both the even and
odd-states, the solid lines are fitted exponentials. Inset:
Switching between even and odd-states observed in non-averaged
measurements at $n_g=1$ with an incident rf-carrier power of -108
dBm. (b) The time constants, deduced from the previous histogram, as
a function of the rf-carrier power. }\label{}
\end{figure}

State parity is ascertained by comparing $V_{IF}$ to a threshold
half-way between the even and odd levels. The distribution of times
spent in the even and odd-states is measured and plotted in a
histogram (fig. 3(a)). We fit an exponential decay $e^{-t/t_{j}}$ to
the histogram, using the parameter $t_{j}$ to define the even and
odd-state lifetimes. For the data in fig. 3(a) $t_{odd}=35$ $\mu$s
and $t_{even}=110$ $\mu$s. The data is well-fitted indicating that
the tunnel processes obey Poissonian statistics. A recent study of
two-level systems has shown that finite receiver bandwidth can have
a significant effect on time constants resulting from an analysis of
this type \cite{naa05}. The majority of time constants measured are
$>10\mu$s and our receiver bandwidth is 1 MHz hence, using the
analysis in \cite{naa05}, we expect the resulting systematic error
to be $<10\%$.

We measure the even and odd-state lifetimes as a function of
rf-carrier power to investigate the effect of the rf-measurement
(fig. 3(b)). A strong reduction in $t_{odd}$ and only a slight
reduction in $t_{even}$ is noticed. The likely cause of reduction of
the odd-state lifetime is that the rf-carrier causes heating which
thermally activates a quasiparticle off the island. The small
reduction in $t_{even}$ indicates that the rf-carrier doesn't
significantly increase the quasiparticle population in the leads,
and any increase in temperature doesn't strongly affect the
poisoning rate.

With the aim of determining the thermal activation of the poisoning
and un-poisoning events we study the even and odd-state lifetimes as
a function of temperature. A low rf-power (-112 dBm) was chosen to
minimize heating by the carrier signal. A reduction in $t_{odd}$
occurs as the temperature is increased, which agrees with the
measurements in \cite{naa06} and is due to thermal excitation of
quasiparticles out of the quasiparticle trap formed on the island
(fig. 4(a)). Considering the free energy change of this transition,
we expect the time constant to be approximately
$t_{odd}=\frac{e^2R}{2}\frac{exp(\delta E/kT)-1}{\delta E}$
\cite{gandd}, where $\delta E$ is the quasiparticle trap depth, and
R the average tunnel junction resistance. Fitting the data
\cite{comment1} to this thermal activation model we find an
experimental value of $\delta E=50\pm4$ $\mu$eV for the trap depth.
The expected quasiparticle trap depth can be determined by
considering the energy difference between the poisoned and
un-poisoned states \cite{aum04}. At the supercurrent peak ($n_g=1$)
the maximum unpoisoned energy (corresponding to the excited-state,
and assuming no superconducting phase difference across the device)
is $E_u=E_c+\frac{E_J}{2}$, while the poisoned state energy has a
minimum energy of $E_p=\delta\Delta=\Delta_i-\Delta_l$. It is useful
to note that trap depth changes with gate bias ($n_g$) and an
investigation of this was carried out in \cite{naa06}. For our
analysis we use the values of $E_c=77$ $\mu$eV and $E_J=33$ $\mu$eV
measured for the SCPT and $\delta\Delta=42\pm18$ $\mu$eV as measured
from the SIS junctions. Calculating $\delta E=E_u-E_p$ a
quasiparticle trap with depth $52\pm18$ $\mu$eV is expected, hence
close agreement to the experimental value is shown.

A constant $t_{even}$ is measured up to $T\sim180$ mK, with the
value being reduced by thermal activation at higher temperatures.
The poisoning rate is expected to depend linearly on both the tunnel
barrier conductance $G_i$ (taken to be in units of $\frac{e^2}{h}$)
and the density of quasiparticles in the leads. An expression for
the poisoning rate was deduced in \cite{lut05} to be
$t_{even}^{-1}\approx AT^{\frac{1}{2}}\exp(-\Delta_l/kT)$ for the
case where temperature is small compared to trap depth ($kT\ll\delta
E$) and
$A=\frac{G_1+G_2}{4\hbar}\sqrt{\frac{k\Delta_l}{2\pi}}\sqrt{\frac{E_u-E_p}{\Delta_l+\Delta_i+E_u}}$.
The temperature dependence of the poisoning rate can be explained by
a constant low temperature quasiparticle population (causing rate
$t_{c}^{-1}$) and, at higher temperatures, the presence of thermally
excited quasiparticles. Adding the rates due to these two
populations we obtain
$t_{even}^{-1}=t_{c}^{-1}+AT^{\frac{1}{2}}\exp(-\Delta_l/kT)$.
Figure 4(a) shows a fit to the data, with all parameters free, and
shows good agreement with some deviation above 300 mK. The
parameters we deduce from the fit are a constant low-temperature
even-state lifetime of $t_{c}=102\pm2$ $\mu$s,
$A=1.8\pm0.2\times10^{20}$ $K^{-0.5}s^{-1}$ and $\Delta_l=213\pm31$
$\mu$eV. The value of $\Delta_l$ is in close agreement with the
measurements from the SIS junctions. The calculated value of A is
$0.92\times10^{20}$ $K^{-0.5}s^{-1}$, hence we see approximate
agreement between the theory and the temperature dependent
experimental poisoning rate. Poisoning rate (slightly) increases
with trap depth in this regime and the theoretical underestimate may
be related to the greater trap depth indicated by measurement of
$t_{odd}$.

\begin{figure}
\includegraphics[width=6.8cm]{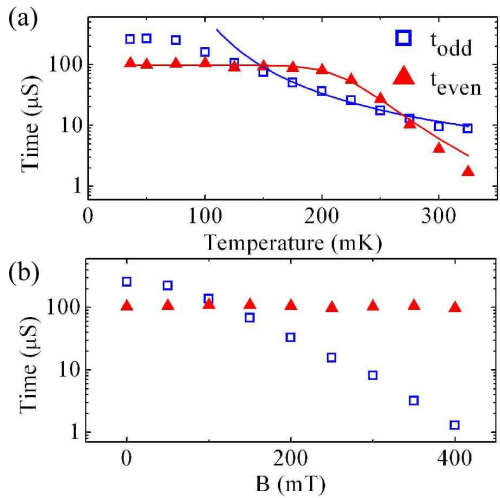}\\
\caption{(a) The time constants $t_{even}$ and $t_{odd}$ determined
as a function of temperature with an rf-carrier power of -112 dBm.
The fit to $t_{odd}$ is from a model of thermal activation of
quasiparticles off the island. For $t_{even}$, the fitted line
includes both a constant quasiparticle poisoning rate and thermal
activation of quasiparticles across the superconducting gap. (b)
Even and odd lifetimes as a function of magnetic field.}\label{}
\end{figure}

We are able to define an effective quasiparticle temperature
($T_{qp}$). This is the temperature that causes a quasiparticle
tunneling rate, due to the thermal excitation of quasiparticles in
the leads, equal to the constant low-temperature value
($t_{c}^{-1}=AT_{qp}^{\frac{1}{2}}exp(-\Delta_l/kT_{qp})$). For this
device $T_{qp}=228$ mK, which is slightly greater than the electron
temperature ($T_{e}\sim150$ mK) as estimated from the fit to
$t_{odd}$. Due to the relatively long recombination time of
quasiparticles ($1-10$ $\mu$s) \cite{lev68}, quasiparticles created
by microwave radiation in the leads can cause $T_{qp}>T_e$. For
previous measurements (on devices with un-engineered $\Delta$) in
which a lack of poisoning was observed e.g. \cite{joy94}, either a
low quasiparticle temperature was achieved or measurement bandwidth
was insufficient to resolve poisoning events.

We also perform a quantitative study of the even and odd state
lifetimes in the presence of an in-plane magnetic field, noting that
magnetic fields have previously been used to change the periodicity
of CPBs \cite{gun04}. Little change is noticed in $t_{even}$ as the
field is increased (fig. 4(b)). Presumably the poisoning rate
remains approximately constant until the quasiparticle trap becomes
a barrier and quasiparticles have to be thermally excited onto the
island. However, there is a large reduction of $t_{odd}$, indicating
that the quasiparticle trap becomes shallower and quasiparticles can
more easily be thermally excited out. For thinner films the critical
field increases \cite{mes71}, indicating a greater reduction in
$\Delta_l$ than $\Delta_i$ (effectively increasing $\delta\Delta$)
at a finite fields. For example, performing a fit to a temperature
dependence of $t_{odd}$ at $B=150$ mT, we find a reduced value of
$\delta E=27\pm2$ $\mu$eV.

In summary, we employed the change in $\Delta$ with aluminum film
thickness to fabricate a SCPT with a reduced quasiparticle trap on
the island. Individual quasiparticle poisoning events were measured
and the resulting statistics analyzed to determine time constants
for the even and odd-state lifetimes. An important aspect of the
experiment was measurement of the thermal activation of poisoning
rate. This enabled a quasiparticle temperature to be determined
which will be a useful parameter to compare different devices and
experimental setups. Furthermore, we expect that by reducing the
island film thickness (increasing $\delta\Delta$) to create a
quasiparticle barrier on the island, or by introducing quasiparticle
traps (therefore increasing $t_{even}$), we will be able to
fabricate SCPTs and CPBs with negligible quasiparticle poisoning.

The authors would like to thank A. C. Doherty, G. J. Milburn, O.
Naaman, J. Aumentado, R. Lutchyn and D. Reilly for helpful
discussions and D. Barber and R. P. Starrett for technical support.
This work is supported by the Australian Research Council, the
Australian government and by the US National Security Agency (NSA)
and US Army Research Office (ARO) under contract number
DAAD19-01-1-0653.


\begin{thebibliography}{19}
\bibitem[{\citenamefont{eq}(2004)\citenamefont{eq}}]{comment1}
\bibnamefont{Assuming saturation of the electron temperature and heating effects
associated with the rf-carrier we only fit the higher temperature
data-points ($T\geq 150$ mK).}
\end{thebibliography}
\end{document}